\documentclass[aps,twocolumn,floatfix,showpacs,amssymb]{revtex4}
\usepackage{color}
\usepackage{hyperref}
\usepackage[utf8]{inputenc}
\usepackage{graphicx,bm}
\usepackage{amsmath}
\usepackage{braket}

\newcommand{\expos}{s}
\newcommand{\Radius}{R}
\newcommand{\Rsep}{R_{\rm sep}}

\newcommand{\Erad}{E_R}
\begin{document}
\title{Universality of isolated $N$-body resonances at large scattering length}
\author{Ludovic Pricoupenko}
\affiliation
{
Laboratoire de Physique Th\'{e}orique de la Mati\`{e}re Condens\'{e}e, Sorbonne Universit\'{e},  CNRS UMR 7600, F-75005, Paris, France.
}
\date{\today}

\begin{abstract}
Non Efimovian $N$-body resonances are investigated in the regime of a large two-body s wave scattering length.  In view of a universal description of low-energy bound and quasi-bound states, a contact model is introduced. The modeling requires two parameters in addition to the scattering length. Using a modified scalar product, the contact model provides a normalization of bound states, possibly not square-integrable, that coincides with that of the corresponding finite range model. 
\end{abstract}
\date{\today}
\maketitle

The regime in which $N$-body quantum systems exhibit zero-energy resonances or shallow states is of great interest in various fields, from hadronic and nuclear physics to condensed matter and ultracold atoms. An interesting class of these phenomena, which can be called $N$-body resonance for short, occurs when the two-body s-wave scattering length ($a_{\rm sc}$) between a fraction of the pairwise interacting particles is large relative to the characteristic range of the interactions ($R_{\rm p}$)\cite{Ham17,Nai17,Son22,Nai22,Nai23,Hiya22,Hon22}. 
This scale separation makes possible the existence of shallow bound states of large extension with respect to the microscopic details of the interaction and it is expected that universal laws can be found in this regime. For more than two particles, the Efimov effect is a fascinating example of such a regime which is predicted up to five particles and is studied experimentally at the three-body level in ultracold atoms by using the Feschbach mechanism \cite{Efi70,Cas10,Baz17}. The reference Hamiltonians that describe this variety of systems are characterized by short range potentials where the small distance details depend on the particular physics involved (pion exchange in nuclear physics, van der Waals and hard core interaction in atomic physics \dots).

In this context, to emphasize the universal character of $N$-body resonances, it is intellectually appealing to replace the finite range interactions of the reference model by pure contact interactions. In this approach, the wave function is a solution of the free Schr\"{o}dinger equation almost everywhere with possibly external potentials, while all particle interactions are replaced by specific singularities at the contact of two or more particles. A remarkable feature is that in this regime, the $N$-body problem for contact interactions can be reduced to the study of a hyper-radial two-dimensional (2D) Schr\"{o}dinger equation with an effective long-range inverse square potential of the form ${\expos^2/\rho^2}$ for ${\rho \ll |a_{\rm sc}|}$. The Efimov effect occurs for imaginary values of ${\expos}$, where this potential is attractive, regardless of the microscopic details of the short-range interactions in the reference model. This regime has been studied in depth \cite{Nai17}. In what follows, one is interested in the non-Efimovian regime where ${\expos}$ is real. In this regime, the $N$-body resonances do not have the same nature because of the repulsive hyper-radial barrier. They occur only for a fine tuning of the finite range  interactions in the reference model and are therefore isolated resonances \cite{Nis08}. Despite the success of the universal Efimov theory, the development of the contact model for $N$-body resonances has been slowed down in the non-Efimovian domain by what can be called the  'normalization catastrophe', i.e. the fact that for ${\expos \ge 1}$, the contact bound states are not square integrable. This serious drawback may give the impression that $N$-body resonances in this regime do not have any universal character \cite{Pet03,Nis08,Saf13,Gao15,Cor15,Kar16}. Nevertheless, the same problem was solved in the contact model for a 3D two-body system with a high partial wave resonance by introducing the notion of modified scalar product \cite{Pri06a,Pri06b}. This suggests that it is possible to find an analogous method here \cite{Note_Felix,Wer06b,Wer08}.

The main results of this paper are as follows: ${(i)}$ Low energy $N$-body isolated resonances are universally described with three parameters: the two-body scattering length ${a_{\rm sc}}$, an effective radius ${\Radius}$ and a detuning parameter ${\epsilon}$; ${(ii)}$ At unitarity (${|a_{\rm sc}|=\infty}$), for negative detuning ${\epsilon<0}$ a single bound state always exists whereas for a positive detuning ${\epsilon>0}$, a long-lived quasi-bound state exists only for ${\expos\ge1}$; ${(iii)}$ The states can be described using a contact model where the interactions are replaced by two- and $N$-body contact conditions; ${(iv)}$ The equivalence between the contact model and the reference model is finalized by introducing a modified scalar product. For shallow bound states, even when they are not square-integrable (i.e. for ${\expos \ge 1}$), this metric gives the same norm as that obtained in the reference model; ${(v)}$ The modified scalar product gives an upper bound for the effective radius parameter of the contact model. All the results can be qualitatively understood through a mapping to a 3D two-body $l$-wave resonant system, which will be denoted as the '3D mapping'.

\begin{figure}
\includegraphics[width=8.5cm]{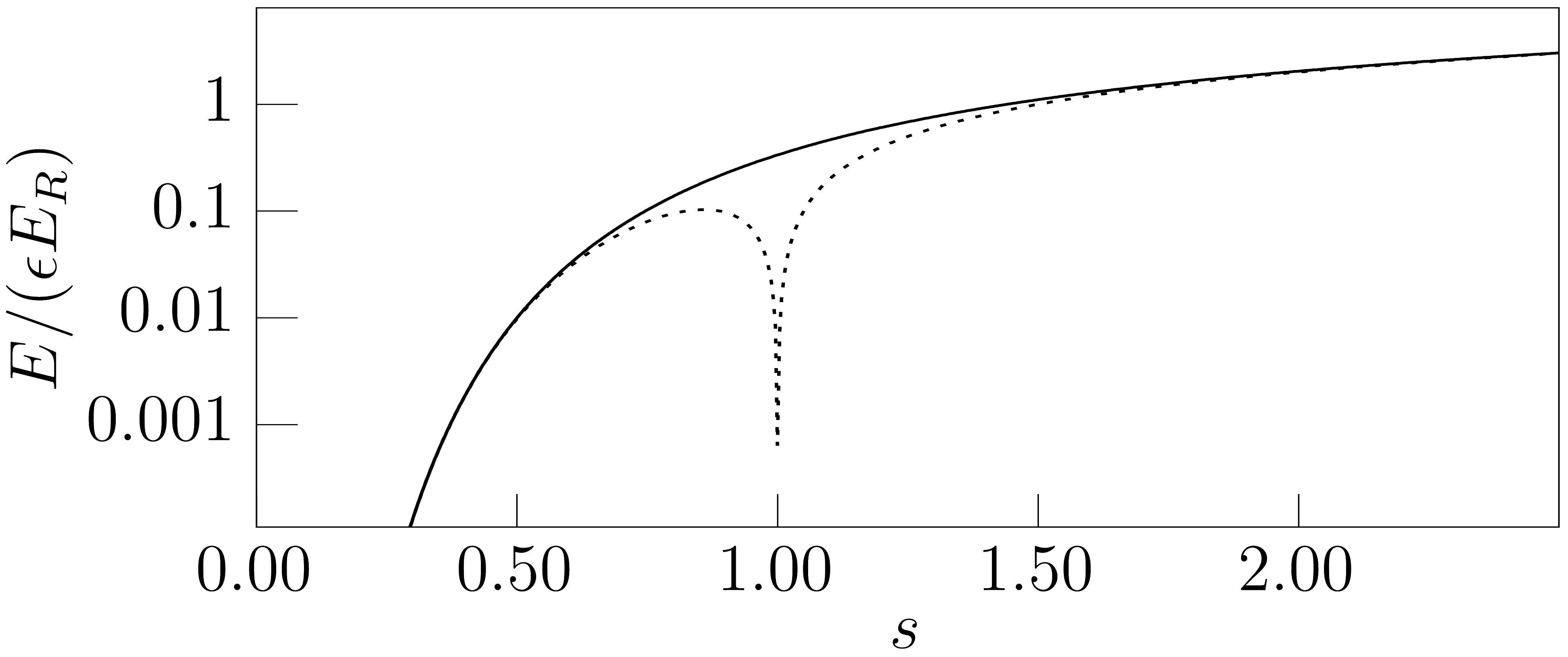}
\caption{Solid line: Ratio ${E/(\epsilon \Erad)}$ ${(\epsilon<0)}$ of the binding energy to the detuning  deduced from Eq.~\eqref{eq:CL} as a function of ${\expos}$ for ${\epsilon=-10^{-2}}$. Dotted line: limit solutions given by Eq.~(\ref{eq:Bound_small_s}) for ${\expos<1}$ and Eq.~(\ref{eq:Bound_lim}) for ${\expos>1}$.}
\label{fig:sol}
\end{figure}

One considers $N$ particles labelled by $i$, of mass ${m_i}$ and spatial coordinates ${\mathbf r_i}$, in their center of mass frame. The choice made for the definitions of the hyper-radius vector ${\boldsymbol \rho}$, hyper-radius ${\rho}$ and of the hyper-angle ${\Omega}$ for a given reference mass ${m_{\rm r}}$ is detailed in the supplemental material. In what follows, the contact model is constructed in four steps. First, one defines what is the contact state ${|\Psi \rangle}$ associated with a given reference state ${|\Psi_{\rm ref}\rangle}$ of the reference model. For this purpose, the reference state is divided into an inner state ${|\Psi_< \rangle}$ and an outer state ${|\Psi_> \rangle}$~:
\begin{equation}
|\Psi_{\rm ref} \rangle = | \Psi_< \rangle + | \Psi_> \rangle .
\label{eq:decomposition}
\end{equation}
The outer state lies in the outer domain ${\mathcal D_{\boldsymbol \rho}}$ where none of the pair $(ij)$ of interacting particles is in the potential range ${(r_{ij}<R_p)}$, so that ${\langle \rho, \Omega | \Psi_> \rangle = 0}$ in the complementary domain of ${\mathcal D_{\boldsymbol \rho}}$. The inner state $| \Psi_< \rangle $ is associated with short distances behavior and cannot be described by the contact model: ${\langle \rho, \Omega | \Psi_< \rangle = 0}$ in the outer domain. By definition, the contact state ${|\Psi \rangle}$ approximates the external state in the outer domain $\mathcal D_{\boldsymbol \rho}$:  ${\langle \rho, \Omega | \Psi \rangle\simeq \langle \rho, \Omega | \Psi_> \rangle}$ and is also a solution of the free Schr\"{o}dinger equation everywhere except at the contact of two or more interacting particles. Second, s wave resonant two-body interactions are replaced by contact conditions. For  a pair, say $(12)$, interacting resonantly in an s wave the two-body contact condition for $r_{12}\to 0$ is  
\begin{equation}
\langle \mathbf r_1 \dots \mathbf r_N | \Psi \rangle \propto \left(\frac{1}{a_{\rm sc}} -\frac{1}{r_{12}}  \right) + O(r_{12}) .
\label{eq:contact_ij}
\end{equation}
Two-body contact conditions for the other interacting pairs are defined in the same manner. Third, one considers the region ${\rho \ll |a_{\rm sc}|}$ where the contact state behaves as in the unitary limit ${|a_{\rm sc}|=\infty}$. In this region, the set of equations deduced from the two-body contact condition of each interacting pair is scale invariant in the hyper-radius. The contact state is thus separable in the hyper-radius and the hyper-angles:
\begin{equation}
\langle \Omega, \rho |\Psi \rangle = \rho^{\frac{5-3N}{2}} F(\rho)  \Phi(\Omega) ,
\label{eq:separability}
\end{equation}
where ${\Phi(\Omega)}$ is a normalized eigenstate (${\langle \Phi |\Phi\rangle=1}$) of the Laplacian $\Delta_{\Omega}$ on the unit sphere: 
${
\Delta_\Omega \Phi(\Omega)=-\Lambda \Phi(\Omega)
}$, 
with the boundary conditions obtained by the set of the two-body contact conditions of the form given by Eq.\eqref{eq:contact_ij}. One then introduces the notion of 'separability region' where the reference state is also separable. This region is defined by the spatial domain ${\mathcal D_{\boldsymbol \rho}}$ and the condition ${\rho \ll |a_{\rm sc}|}$. There exists a minimal radius $\Rsep$ of the order of $R_{\rm p}$  such that ${\Rsep < \rho \ll |a_{\rm sc}|}$ in the separability region. By construction, the contact and reference radial functions (almost) coincide in the separable region. For ${0<\rho \ll |a_{\rm sc}|}$, the radial function ${F(\rho)}$ satisfies~:
 \begin{equation}
-\frac{\hbar^2}{2m_{\rm r} }  \left(\partial_\rho^2 +\frac{1}{\rho} \partial_\rho \right)F(\rho)   + \frac{\hbar^2 \expos^2}{2m_{\rm r}  \rho^2} F(\rho) = E F(\rho)
\label{eq:Schrodi_hyper_radial}
\end{equation}
where ${\expos^2=\Lambda+\left(\frac{3N-5}{2}\right)^2}$.  At unitarity, Eq.~\eqref{eq:Schrodi_hyper_radial} is formally equivalent to the radial equation of a 2D two-body problem with an angular momentum in the continuum (i.e. classical). The form of the $N$-body contact condition shared by all the contact wave functions will be deduced from the short distance behavior of the radial function in the unitary limit. One therefore focus on this regime where Eq.~\eqref{eq:Schrodi_hyper_radial} is always valid for ${\rho>0}$. The bound state solutions of Eq.~\eqref{eq:Schrodi_hyper_radial}  are given by the Macdonald function
\begin{equation}
F(\rho)=\mathcal A K_{\expos}(q \rho) 
\label{eq:F_unitary}
\end{equation}
where ${\mathcal A}$ is the normalization constant. If there is no boundary condition at ${\rho=0}$, all values of the binding wave number are possible, which is a physical nonsense. Following Efimov's seminal paper, the necessary contact condition can be deduced by imposing a specific value on the log-derivative of the contact wave function considered at the effective radius $\Radius$ \cite{Efi71,Log-deriv,nodal}:
\begin{equation}
\frac{\partial_\rho F(\rho)}{F(\rho)}\biggr|_{\rho=\Radius} = \frac{\epsilon-\expos}{\Radius}.
\label{eq:log_derivative}
\end{equation}
From Eq.~\eqref{eq:F_unitary}, a zero-energy $N$-body resonance occurs for a vanishing detuning parameter $\epsilon$ \cite{s=.5}. The effective radius sets a high energy scale $\Erad$ given by :
\begin{equation}
\Erad = \frac{\hbar^2}{2 m_{\rm r} \Radius^2} .
\label{eq:high_energy}
\end{equation}
As in the Efimov's theory at unitarity, the log-derivative condition breaks the scale invariance related to the ${1/\rho^2}$ potential and gives rise to a quantum anomaly \cite{Cam03}. The 3D mapping corresponds to the equivalence of Eq.~\eqref{eq:Schrodi_hyper_radial} with a 3D radial equation for a $l$-wave state obtained through the change of function $F(\rho)=\sqrt{\rho} f(\rho)$ which gives ${\expos=l-\frac{1}{2}}$ and corresponds also to a $l$-wave resonant problem when one imposes the log-derivative condition of Eq.~\eqref{eq:log_derivative}. The parameters $(\epsilon,\Radius)$ are thus related to the two parameters in the low energy  $l$-wave resonant scattering amplitude \cite{Lan81,Pri06b} (for non specialists, details are given in the supplemental material). Before deducing the contact condition, it is interesting to describe the low energy solutions (i.e. ${|E|\ll \Erad}$ or ${|\epsilon|\ll1}$) deduced from Eq.~\eqref{eq:log_derivative}. For this purpose, one uses the truncated expansion of the radial wave function in Eq.~\eqref{eq:F_unitary}, considered as a function of the variable ${z=q\rho}$ for ${z\to 0}$ \cite{seriesKs}. This series must at least include the terms $z^{\pm\expos}$ which are the two possible zero energy solutions of Eq.~\eqref{eq:Schrodi_hyper_radial} for $z\to0$. One obtains the equation verified by the energy $E$ of a shallow state:
\begin{equation}
\sum_{k=0}^{k_{\rm max}}  a(k,\expos,\epsilon) \left(\frac{E}{\Erad}\right)^k   =\frac{\pi}{\sin (\pi s)} \left(\frac{-E}{\Erad}\right)^{\expos}
\label{eq:CL}
\end{equation}
where 
\begin{equation}
a(k,\expos,\epsilon )=4^{\expos-k} \left(\frac{2k-\epsilon}{2\expos-\epsilon}\right) \frac{\Gamma(\expos+1) \Gamma(\expos-k)}{k !} .
\label{eq:coeff_a}
\end{equation}
The cut-off ${k_{\rm max}}$ in Eq.~\eqref{eq:CL} has to be carefully chosen. Indeed, the term of order ${2k\pm \expos}$ in the expansion of the modified Bessel function is proportional to the factor ${\Gamma(\mp\expos-k)}$, which provides an anomalously large contribution when ${\mp\expos-k}$ is in the vicinity of zero or of a negative integer value \cite{seriesKs}. When truncated, this expansion can thus lead to a very bad approximation. Nevertheless, one can verify that as ${\expos}$ tends to ${n}$ for a fixed value of ${z}$, the spurious singularity of the term in ${z^\expos}$ is compensated by the one of the term in ${z^{-\expos+2n}}$ \cite{compensation}. To avoid also the next order spurious singularity when ${n<\expos}$ and ${\expos \simeq n}$, one introduces the small positive number $\eta$ and the cut-off ${k_{\rm max}}$ is chosen as ${k_{\rm max} = \lceil \expos - \eta \rceil}$
with typically $\eta \simeq 0.2$. For small and negative values of the detuning (i.e. ${\epsilon<0}$), a unique shallow bound state is always found. Near the threshold of the Efimov regime, i.e. in the limit  ${0<\expos\ll 1}$, the binding energy is:
\begin{equation}
E \simeq - 4 \Erad \left[\frac{ -\epsilon  \expos \sin(\pi \expos)
 \Gamma(\expos)^2 }{(2\expos - \epsilon) \pi} \right]^{1\over \expos} .
\label{eq:Bound_small_s}
\end{equation}
For ${\expos=0}$, ${E=-4 \Erad \exp\left(2/\epsilon-2\gamma_E\right)}$ where ${\gamma_E}$ is the Euler's constant. For increasing values of ${\expos}$, the  first order term ${k=1}$ in  Eq.~\eqref{eq:CL} cannot be neglected and the truncation of the sum at this order gives the effective range approximation in the 3D mapping. In the limit of large values of ${\expos}$ and  for a small detuning ${|\epsilon| \ll1}$, summation of the terms ${k\ge 2}$ in Eq.~\eqref{eq:CL} provides a negligible contribution. Consequently, one can safely neglects these terms and find the limit solution for ${\expos \gg 1}$: 
\begin{equation}
E \simeq 2(\expos-1) \epsilon \Erad .
\label{eq:Bound_lim}
\end{equation}
For a small and positive detuning (i.e. ${0<\epsilon\ll1}$) the solutions of Eq.~\eqref{eq:CL} are complex with a positive real part and  an imaginary part that can be chosen negative: ${E=E_{\rm r} - i \Gamma/2}$. Long-lived quasi-bound states defined by a vanishing ratio ${\Gamma/E_r}$ are found for ${\expos \ge 1}$. At the threshold ${\expos=1}$, one finds  in the limit $\epsilon \to 0$ \cite{s=1}:
\begin{equation}
E_{\rm r} \simeq - \frac{2\epsilon \Erad }{W_{-1}(-x)}  \quad , \quad  \frac{\Gamma}{2E_{\rm r}} \simeq \frac{- \pi}{1+\ln x} ,
\label{eq:quasi-bound_neq1}
\end{equation}
where ${x=e^{2\gamma_E}\epsilon/2}$ and ${W_{-1}}$ is the lower branch of the Lambert function. For sufficiently large values of ${\expos}$, the quasi-bound state energy is given by
\begin{equation}
E_{\rm r} \simeq 2(\expos-1) \epsilon \Erad \quad , \quad  \frac{\Gamma}{E_r} \simeq 
\frac{4^{1-s}\pi(\expos-1)}{[\Gamma(s)]^2} \left(\frac{E_r}{\Erad}\right)^{\expos-1}
\label{eq:quasi-bound}
\end{equation}
The existence of this quasi-bound state is due to the repulsive ${\expos^2/\rho^2}$ barrier, analogous to the centrifugal barrier obtained in the 3D mapping for high partial waves. In the limit of large values of ${\expos}$, the effective centrifugal barrier grows, explaining the reason why Eq.~(\ref{eq:quasi-bound}) predicts very long lived quasi-bound states with a ratio ${\Gamma}/{E_{\rm r}}$ that tends to zero. On the contrary for ${\expos<1}$, the barrier is not strong enough to support a long-lived quasi-bound state. It is interesting to compare the solution of Eq.~\eqref{eq:CL} with the limit solutions given by Eq.~\eqref{eq:Bound_small_s} for ${\expos<1}$ and Eq.\eqref{eq:Bound_lim} for ${\expos>1}$. This is done in Fig.~(\ref{fig:sol}) with the plot of ${E/(\epsilon \Erad)}$ for ${\epsilon=-0.01}$. The approximation in Eq.~\eqref{eq:Bound_small_s} fails for values of ${\expos}$ greater than about ${.7}$, a number compatible with the fact that in the 3D mapping, ${\expos=1/2}$ corresponds to a s wave resonance with a single parameter model \cite{s=.5}. Interestingly, the approximation for large values of ${\expos}$ in Eq.~\eqref{eq:Bound_lim} becomes relevant  for ${\expos}$ higher than but of the order of unity. Again this property is well understood with the 3D mapping where the binding energy of a shallow state for ${l \ge1}$ is given accurately in the effective range approximation. For a vanishing value of $\epsilon$, the spectrum tends to the limit solutions except at ${\expos=1}$.

The $N$-body contact condition is such that it leads to the solutions of Eq.~\eqref{eq:log_derivative} in the low energy limit. It corresponds to imposing a specific linear combination of the coefficients of the series of the radial wave function $F(\rho)$ for a vanishing value of the hyper-radius ${\rho\to 0}$. For this purpose, it is convenient to introduce the operator
\begin{equation}
\lim_{\rho\to 0} \bigl]\rho^\expos , F(\rho)\bigr[ 
\label{eq:extraction}
\end{equation}
which gives the coefficient of the term $\rho^\expos$ in this series. The contact condition is obtained by establishing a mapping between the first terms of the series and the condition in Eq.~\eqref{eq:CL}. Using the behavior of the Macdonald function ${K_\expos(z)}$ when ${z\to 0}$, one finds \cite{seriesKs}:
\begin{equation}
\lim_{\rho\to 0} \bigg] \Radius^\expos \rho^\expos + \sum_{k=0}^{k_{\rm max}}\left(\frac{2k-\epsilon}{2\expos-\epsilon}\right) (\Radius\rho)^{2k-\expos},F(\rho)\bigg[  =0 .
\label{eq:contact_N_body}
\end{equation}
This contact condition  which  in the low energy limit, is equivalent to the log-derivative condition \eqref{eq:log_derivative}, can be used for any contact wave function in the regime of large scattering length (${|a_{\rm sc}| \gg R_p}$) \cite{Contact_Efimov}. For integer values of ${\expos=n}$, the contact model of a $N$-body isolated resonance is formally equivalent to a contact model for the 2D two-body problem with a resonant interaction in the $n^{\rm th}$ partial wave. In this case, the series of $F(\rho)$ contains terms of the form $\rho^n \ln(\rho q c_n)$ \cite{seriesKn}. For ${n\ge1}$, the contact condition in Eq.~\eqref{eq:contact_N_body} is then modified by taking ${k_{\rm max}= n-1}$ and by defining the action of the operator in Eq~\eqref{eq:extraction} for such terms as \cite{seriesKn,s=0}:
\begin{equation}
\lim_{\rho\to 0} \bigl]\rho^n , \rho^n \ln ( \alpha \rho)\bigr[= \ln (\alpha \Radius) +1/(2n-\epsilon).
\end{equation}

In the domain defined by Eq.~\eqref{eq:log_derivative}, two contact states of different energies are not mutually orthogonal and a bound state is not normalizable in the usual sense for ${\expos \ge 1}$. This problem is linked to the behavior of the contact radial function for ${\rho<\Rsep}$ and it is tempting to introduce a subtraction or a cut-off in the radial integral to avoid the unphysical divergence. As expected from the 3D mapping, the same problem exists in a contact model of a resonant interaction with $l\ge1$ and the 'normalization catastrophe' is solved by introducing a modified scalar product. The issue is to do again this type of regularization in an equivalent rigorous manner. To extend the method to the present situation, one can notice that at unitarity and using  Eq.~(\ref{eq:Schrodi_hyper_radial}), two contact states of radial functions ${F(\rho,E)},{F(\rho,E')}$, with different energies ${E' \ne E}$, verify 
\begin{multline}
\int_R^\infty \rho F(\rho,E')^* F(\rho,E) d\rho  = \frac{\hbar^2\Radius}{2m_{\rm r}(E'-E)}\\
\times \left[ F(\Radius,E) \partial_\Radius F(\Radius,E')^* -  (E\leftrightarrow E')^*\right]
\label{eq:orthogonality}
\end{multline}
Using Eq.~(\ref{eq:log_derivative}) in the right hand side of Eq.~\eqref{eq:orthogonality} proves that the integral is identically zero. It is thus natural to introduce the following modified scalar product in the center of mass frame
\begin{equation}
\left( \Psi'|\Psi \right)_0 \equiv  \int_{\rho>R}d^{3N-3}\mu\langle \Psi'|\Omega,\rho \rangle \langle \Omega, \rho |\Psi \rangle ,
\label{eq:modified_scal_prod}
\end{equation}
where   ${d^{3N-3}\mu}$ is the measure of integration in the $3N-3$ dimensional space. In the domain of functions defined by Eq.~\eqref{eq:log_derivative}, one finds that using  the modified scalar product, the contact model is a self-adjoint extension of the $N$-body kinetic operator $H_0$ \cite{Ara04}:
\begin{equation}
\left( \Psi'|H_0 \Psi \right)_0 = \left( H_0\Psi' | \Psi \right)_0 
\label{eq:self_adjoint}
\end{equation}
where the surface terms in the hyper-radial integration have been eliminated thanks to Eq.~\eqref{eq:log_derivative}. Using the self-adjoint character of the reference Hamiltonian and the log-derivative condition in Eq.~\eqref{eq:log_derivative}, one obtains the following crucial property: the normalization of a contact bound state obtained from the  modified scalar product  ${\left( \Psi|\Psi \right)_0=1}$,  coincides with the normalization of the associated reference state ${|\Psi_{\rm ref}\rangle}$ where one uses the standard scalar product ${\langle\Psi_{\rm ref}|\Psi_{\rm ref}\rangle=1}$ (the detailed proof is given in the supplemental material). Hence, in the low energy limit, for two contact states  ${|\Psi\rangle, |\Psi'\rangle}$ associated with the reference states  ${|\Psi_{\rm ref}\rangle, |\Psi'_{\rm ref}\rangle}$:
\begin{equation}
\left( \Psi' |\Psi \right)_0 = \langle \Psi'_{\rm ref} |\Psi_{\rm ref}\rangle  .
\end{equation}
Using the decomposition of the reference state in Eq.~\eqref{eq:decomposition}, one finds the norm of the inner state
\begin{equation}
\langle \Psi_<(E)|\Psi_<(E) \rangle = |{\mathcal A}(E)|^2 \int_{\Radius}^{\Rsep} \rho K_{\expos}(q\rho)^2 d\rho 
\label{eq:inner_prod_scal}
\end{equation}
where ${\mathcal A(E)}$ is the normalization factor in the separable region and the contact wave function is given by Eqs.~(\ref{eq:separability},\ref{eq:F_unitary}). Two important remarks are then in order. First, the positivity of the norm implies from Eq.~\eqref{eq:inner_prod_scal}
\begin{equation}
 \Radius < \Rsep   .
\label{eq:radius_inequality}
\end{equation}
Using the 3D mapping, this last inequality is reminiscent of the Wigner bound obtained for high partial waves, a result that was also obtained in this context by using a modified scalar product \cite{Pri06a,Pri06b,Ham10}. Second, the expression in Eq.~\eqref{eq:inner_prod_scal}  has been derived in the unitary limit. However, it is only related to the behavior of the reference state in the separable region and thus it remains valid for finite but large values of the scattering length ${(|a_{\rm sc}|\gg \Rsep)}$. On the other hand, the contribution of the external state in Eq.~\eqref{eq:modified_scal_prod} is obtained using the standard scalar product. Therefore, the expression for the modified scalar product in Eq.~\eqref{eq:modified_scal_prod} is also valid when the two-body scattering length is large.  In the interval ${0<\expos<1}$ where the contact state is square-integrable, it is interesting to consider the ratio of the modified norm to the usual norm  
\begin{equation}
r=\frac{\left( \Psi |\Psi \right)_0}{\langle \Psi |\Psi \rangle}
 =\frac{\int_{qR}^{\infty} K_s(z)^2 z dz}{\int_{0}^{\infty} K_s(z)^2 z dz} .
\end{equation}
This ratio is a decreasing function of ${\expos}$ and ${q\Radius}$ respectively. Notable deviations from unity occur in the vicinity of ${\expos=1}$. For example, for ${q\Radius=10^{-2}}$, one finds ${r=.9}$ at ${\expos \sim 0.72}$, which shows that the norm of the internal part of the reference state cannot be neglected when  ${\expos}$ is close to unity. At the threshold ${q\Radius=0}$, ${r=1}$ for ${\expos<1}$, which means that the system is essentially in the outer domain ${\mathcal{D}_{\boldsymbol \rho}}$ as is the case in a standard 3D two-body s wave resonance. This property is lost for ${\expos\ge 1}$.

At finite values of the two-body scattering length ${a_{\rm sc}}$, the wave function is no longer separable and one is faced with coupled equations on the hyper-angles and the hyper-radius that depend on the nature of the system (number of particles, mass ratios \dots). Interestingly, the case of two identical fermions interacting with an impurity has already been studied by using also a log-derivative condition for the three-body condition and an hyper-spherical expansion \cite{Saf13}. The results of this last paper, can be reinterpreted in the framework of a pure zero-range model, revealing their universal character, including the bound states already found for ${\expos>1}$.

Isolated three-body resonances are predicted for two identical fermions interacting resonantly in the s wave with a sufficiently massive impurity (to avoid the Efimov effect) in Ref.~\cite{Nis08,Saf13,Kar16}. When the two fermions experience a pairwise short range p wave resonant interaction, the shallow 'p-wave induced' states found are particular realizations of isolated three-body resonances~\cite{Nai22}. It is thus possible in principle to achieve tunable isolated $N$-body resonances in heterogeneous two-component Fermi mixtures~\cite{large_s} by combining an optical p wave Feshbach resonance between identical fermions~\cite{Goy10,Yama13} and a magnetic s wave Feshbach resonance between distinct particles. To conclude, the contact model considered at unitarity, can be viewed formally as a two-body 2D or 3D contact model of a resonant $l$-wave where ${l\in \mathbb R^+}$. The results obtained follow from a self-adjoint extension of the Laplacian and of the properties of the inverse radius square potential that may be of interest in other fields.  

\newpage
\begin{widetext}

\section*{Supplemental material}

\section*{A- Definition of the Jacobi and hyper-spherical  coordinates used in the main text}

In this section, the Jacobi variables are introduced for $N$ the particles in the same manner as in Ref.~\cite{Wer06b}. Beginning with the relative coordinates of an interacting pair, say the pair $(12)$, the other Jacobi coordinates are built iteratively by defining at the step $n$ the relative particle formed by the particle $n$ and the relative particle of the step ${n-1}$. One starts by defining the mass $M_j$ and center of mass $\mathbf C_j $ of the set composed of the first $j$ particles:
\begin{equation}
M_j = \sum_{i=1}^j m_i \quad ; \quad  \mathbf C_j = \frac{1}{M_j} \sum_{i=1}^j m_i \mathbf r_i .
\end{equation}
The center of mass of the system is denoted by ${\mathbf C=\mathbf C_N}$. The reduced mass and the coordinates for the relative particle formed by the ${j^{\rm th}}$ particle and the set composed of the first ${j-1}$ particles is
\begin{equation}
\mu_j = \frac{m_{j+1} M_{j}}{M_{j+1}}  \quad ; \quad \boldsymbol \eta_j = \sqrt{\frac{\mu_j}{m_{\rm r}}} \left( \mathbf r_{j+1} - \mathbf C_j \right)  ,
\end{equation}
where ${1 \le j\le N-1}$ and $m_{\rm r}$ is an arbitrary reference mass. The ${N-1}$ vectors ${\{\boldsymbol \eta_1, \boldsymbol \eta_2 \dots \boldsymbol \eta_{N-1}\}}$ form a possible set of Jacobi coordinates. Other sets  of Jacobi coordinates can be defined in the same manner by begining the iteration with another interacting pair. This way, the two-body contact condition for the pair $(ij)$ can be always written in terms of the variable $\eta_1$ of the set of Jacobi coordinates defined from the initial pair $(ij)$. From the coordinates $\{\boldsymbol \eta_i\}$ one defines an ${3N-3}$-dimensional hyper-radius vector $\boldsymbol \rho$,  the hyper-radius $\rho$ 
\begin{equation}
\boldsymbol \rho=\left( \boldsymbol \eta_1, \boldsymbol \eta_2, \dots \boldsymbol \eta_{N-1} \right)  \qquad,\qquad  \rho=\sqrt{\sum_{i=1}^{N-1} \eta_i^2}
\end{equation}
and  the set of angles $\Omega$ parameterized  by the unit vector $\left( \frac{\boldsymbol \eta_1}{\rho}, \dots \frac{\boldsymbol \eta_{N-1}}{\rho} \right)$. In configuration space the degrees of freedom can be then defined by the coordinates $(\mathbf C, \rho, {\Omega})$. In the center of mass frame, the Hamiltonian $H_0$ of the $N$-body system reduces to the kinetic operator 
\begin{equation}
H_0= \sum_{i=1}^N \frac{-\hbar^2}{2m_i} \Delta_{\mathbf r_i} \equiv -\frac{\hbar^2}{2m_{\rm r}} \sum_{i=1}^{N-1} \Delta_{\boldsymbol \eta_i}  \equiv \frac{-\hbar^2}{2m_{\rm r}} \Delta_{\boldsymbol \rho} .
\label{eq:def_kinetic_Nbody}
\end{equation}
It can be expressed in terms of the hyper-radial kinetic operator
\begin{equation}
T_\rho =-\frac{\hbar^2}{2m_{\rm r}}
\left( \partial_\rho^2 + \frac{3N-4}{\rho} \partial_\rho \right)
\end{equation}
and of the Laplacian $\Delta_\Omega$ acting on the hyper-sphere of radius unity:
\begin{equation}
H_0=T_\rho-\frac{\hbar^2}{2m_{\rm r}} \frac{\Delta_{\Omega}}{\rho^2} .
\end{equation}
The expression of $\Delta_\Omega$ is not useful in this work. In the contact model, the stationary Schr\"{o}dinger equation for a state $ |\Psi\rangle$ of energy $E$ can be then written as:
\begin{equation}
\left( T_\rho  -\frac{\hbar^2}{2m_{\rm r}} \frac{\Delta_\Omega}{\rho^2} - E \right) \langle \Omega, \rho |\Psi\rangle = 0 .
\label{eq:Schrodi_hyper_contact}
\end{equation}
This equation is satisfied by the contact state everywhere except at the contact of two interacting particles where Eq.~\eqref{eq:contact_ij} holds or at $\rho=0$ where the $N$-body contact condition is used.

\section*{B- Normalization of the reference state at unitarity}

The following derivation of the normalization of a reference bound state at unitarity is deduced from the log-derivative condition of Eq.~\eqref{eq:log_derivative} which is equivalent to the contact condition in Eq.~\eqref{eq:contact_N_body} in the low energy limit ${|E|\ll \Erad}$. One considers two reference states $\{|\Psi_{\rm ref}(E)\rangle, |\Psi'_{\rm ref}(E')\rangle\}$ of arbitrary energies $E$ and $E'$ in the center of mass frame. From the stationary Schr\"{o}dinger equation,  one obtains :
\begin{equation}
 \langle \Omega, \rho  |\Psi'_{\rm ref}(E')\rangle^* H_0 \langle \Omega, \rho  |\Psi_{\rm ref}(E)\rangle-
 \langle \Omega, \rho  |\Psi_{\rm ref}(E)\rangle H_0 \langle \Omega, \rho  |\Psi'_{\rm ref}(E')\rangle^*
=(E-E') \langle \Omega, \rho  |\Psi'_{\rm ref}(E')\rangle^* \langle \Omega, \rho |\Psi_{\rm ref}(E)\rangle . 
\label{eq:Wronskian_1}
\end{equation}
The operator ${\Delta_\Omega}$ is self-adjoint in the domain of the reference wave functions. Hence, integration of each side of Eq.~\eqref{eq:Wronskian_1} over the unit hyper-sphere and for hyper-radii smaller than a given cut-off ${\rho_M}$, gives
\begin{multline}
\int_{\rho=0}^{\rho=\rho_M} d^{3N-3}\mu\,
\bigl( \langle \Omega, \rho  |\Psi_{\rm ref}'(E')\rangle^* T_\rho \langle \Omega, \rho  |\Psi_{\rm ref}(E)\rangle 
-  \langle \Omega, \rho  |\Psi_{\rm ref}(E)\rangle T_\rho \langle \Omega, \rho |\Psi_{\rm ref}'(E')\rangle^*\bigr)
\\
=(E-E') \int_{\rho=0}^{\rho=\rho_M} d^{3N-3}\mu\, \langle \Omega, \rho  |\Psi_{\rm ref}'(E')\rangle^* \langle \Omega, \rho |\Psi_{\rm ref}(E)\rangle 
\label{eq:Wronskian_2}
\end{multline}
where the measure of integration in the $3N-3$ dimensional space is ${d^{3N-3}\mu = \rho^{3N-4} d\rho d^{3N-4}\Omega}$. For realistic potentials in the reference model, the reference wave function and its radial derivative vanish at the origin $\rho=0$. Equation \eqref{eq:Wronskian_2} can then be transformed into
\begin{equation}
\int_{\rho=0}^{\rho=\rho_M} d^{3N-3}\mu \, \langle \Psi_{\rm ref}'(E')|\Omega,\rho\rangle \langle \Omega, \rho |\Psi_{\rm ref}(E) \rangle
= \frac{\hbar^2\rho_M^{3N-4}}{2m_r} \frac{ 
 \int d^{3N-4} \Omega  W\left[
  \langle \Omega, \rho |\Psi_{\rm ref}'(E')\rangle^*, \langle \Omega, \rho  |\Psi_{\rm ref}(E)\rangle,\rho=\rho_M
\right]}{E'-E} .
\label{eq:Wronskian_3}
\end{equation}
The term  ${W[f,g,\rho=\rho_M]=f \partial_\rho g-g \partial_\rho f}$  in Eq.~\eqref{eq:Wronskian_3} is the Wronskian of the functions $f$ and $g$ with respect to the variable $\rho$, considered at ${\rho=\rho_M}$. For $\rho$ in the separable region  ($\rho>\Rsep$), the reference wave functions are well approximated by their associated contact wave functions:
\begin{equation}
\langle \Omega, \rho |\Psi_{\rm ref}(E)\rangle = \rho^{\frac{5-3N}{2}} F(\rho,E) \Phi(\Omega) 
\qquad ,  \qquad 
\langle \Omega, \rho |\Psi'_{\rm ref}(E')\rangle = \rho^{\frac{5-3N}{2}} F(\rho,E') \Phi'(\Omega) .
\end{equation}
In what follows, one focus on the case where  ${|\Phi'\rangle = |\Phi\rangle}$ and without loss of generality $\langle \Phi| \Phi\rangle=1$ as in Eq.~\eqref{eq:separability}. Then for ${\rho_M>\Rsep}$ Eq.~\eqref{eq:Wronskian_3} transforms  into
\begin{equation}
\int_{\rho=0}^{\rho=\rho_M} d^{3N-3}\mu \langle \Psi_{\rm ref}(E')|\Omega,\rho\rangle \langle \Omega, \rho |\Psi_{\rm ref}(E) \rangle =\frac{\hbar^2\rho_M}{2m_r} \frac{ W\left[  F(\rho,E')^*,F(\rho,E),\rho=\rho_M \right]}{E'-E} .
\label{eq:Wronskian_4}
\end{equation}
Formally, one can consider solutions of the Schr\"{o}dinger equation, $F(\rho,E)$ for arbitrary negative values of $E$. In the unitary limit, the general solution of the hyper-radial problem  is
\begin{equation}
F(\rho,E)={\mathcal A}(E) K_\expos(q \rho) + {\mathcal B}(E) I_\expos(q \rho) ,
\label{eq:continuous_E}
\end{equation}
where ${I_s}$ is the modified Bessel function of the first kind. For each value of the energy $E$, the continuation of  the radial reference function $F(\rho,E)$ for  $\rho<\Rsep$ verifies the log-derivative condition in Eq.~\eqref{eq:log_derivative} and the energy of a bound state is such that ${{\mathcal B}(E)=0}$. Taking the limit ${E' \to E}$ in Eq.~\eqref{eq:Wronskian_4}, where ${E}$ is the energy of a bound state, one finds
\begin{equation}
\int_{\rho=0}^{\rho=\rho_M} d^{3N-3}\mu |\langle \Omega, \rho |\Psi_{\rm ref}(E) \rangle|^2  =  \frac{\hbar^2 \rho_M}{2m_r} 
W\left[\partial_E F(\rho,E)^*,F(\rho,E) ,\rho=\rho_M \right] .
\label{eq:norme_Landau}
\end{equation}
Using the property
\begin{equation}
W[ K_\expos(z),I_\expos(z),z]= \frac{1}{z} ,
\end{equation}
one has from Eq.~\eqref{eq:continuous_E}
\begin{equation}
\frac{\hbar^2 \rho_M}{2m_r} W[\partial_E F(\rho,E)^*,F(\rho,E),\rho=\rho_M]=
-\frac{\rho_M^2 |{\mathcal A(E)}|^2}{2z_M} W\left[z\partial_s K_{\expos},K_{\expos},z=z_M\right]
-\frac{\hbar^2}{2m_r} \partial_E {\mathcal B}(E)^* {\mathcal A}(E) ,
\label{eq:Wronskian_family}
\end{equation}
where ${z_M=q\rho_M}$. Furthermore, using the log-derivative condition Eq.~\eqref{eq:log_derivative}, which is verified by $F(\rho,E)$ and also by $F(\rho,E+dE)$, one finds  
\begin{equation}
\partial_E {\mathcal B}(E)=\frac{m_r {\mathcal A}(E)}{\hbar^2q^2  } \times \frac{z \partial_z( z \partial_zK_\expos(z) )+ (\expos-\epsilon) \partial_zK_\expos(z)}
{  z \partial_zI_\expos(z) +(\expos-\epsilon) I_\expos(z)}\biggr|_{z=z_R} 
\label{eq:partial_B_family}
\end{equation}
where ${z_R=q\Radius}$. In the limit ${z_M\to \infty}$, one has $W\left[z\partial_z K_{\expos},K_{\expos},z=z_M\right]=0$, and thus
\begin{equation}
\langle \Psi_{\rm ref}(E)|\Psi_{\rm ref}(E) \rangle
= \frac{- |{\mathcal A}(E)|^2}{2q^2}\times \frac{z\partial_z (z \partial_z K_\expos(z))+(\expos-\epsilon) z \partial_z K_\expos(z)}{z\partial_zI_\expos(z) +(\expos-\epsilon)I_\expos(z)}\biggr|_{z=z_R} .
\label{eq:prod_tot}
\end{equation}
Using the expression of $\epsilon-\expos$ deduced from Eq.~\eqref{eq:log_derivative}, one obtains a crucial identity for the quantity $J(z_R)$ that appears in the right hand side of Eq.~\eqref{eq:prod_tot}:
\begin{equation}
J(z)\equiv -\frac{z\partial_z (z \partial_z K_\expos(z))+(\expos-\epsilon) z \partial_z K_\expos(z)}{2z\partial_zI_\expos(z) +2(\expos-\epsilon)I_\expos(z)} 
= \frac{z}{2}  W\left[z\partial_z K_{\expos}(z),K_{\expos}(z),z\right]  .
\label{eq:identity_Wronskian}
\end{equation}
In another hand
\begin{equation}
\langle \Psi_>(E)|\Psi_>(E) \rangle = 
\frac{|{\mathcal A}(E)|^2}{q^2} \int_{q\Rsep}^{\infty} u K_{\expos}(u)^2 du
\end{equation}
and from Eq.~\eqref{eq:Wronskian_family} considered at ${\rho_M=\Rsep}$, the norm of the inner state is
\begin{equation}
\langle \Psi_<(E)|\Psi_<(E) \rangle =\frac{|{\mathcal A}(E)|^2}{q^2}
\left[J(q\Radius)-J(q\Rsep)\right] .
\label{eq:norme_int}
\end{equation}
Using the fact that
\begin{equation}
\langle \Psi_{\rm ref}(E) | \Psi_{\rm ref}(E) \rangle
= \frac{|{\mathcal A}(E)|^2}{q^2} J(q\Radius) 
=\langle \Psi_<(E) | \Psi_<(E) \rangle + \langle \Psi_>(E) | \Psi_>(E) \rangle 
\end{equation}
one deduces that 
\begin{equation}
J(z)=\int_z^\infty u K_\expos(u)^2 du 
=\frac{z}{2}\left[z K_{\expos+1}(z)^2 - z K_{\expos}(z)^2 -2 \expos K_{\expos+1}(z)K_{\expos}(z)\right]
\label{eq:exact_integral}
\end{equation}
a result that can be checked numerically at the desired accuracy. Finally, one obtains the desired result at the unitary limit of the two-body interaction~:
\begin{equation}
\langle \Psi_{\rm ref}(E)|\Psi_{\rm ref}(E) \rangle
=|{\mathcal A}(E)|^2 \int_\Radius^\infty \rho K_\expos(q\rho)^2 d\rho . 
\end{equation}

\section*{C- Mapping to a three-dimensional two-body problem with a resonant $l$-wave}

This section give some details about the mapping between the contact model of the isolated $N$-body resonance  at unitarity and the contact model for a two-body problem with a symmetric $l$-wave resonance, denoted in short '3D mapping'. It reminds known results and specifies the relations between $(\epsilon, \Radius)$ and the standard low energy 3D scattering parameter with the notations of Ref.~\cite{Pri06a}.

Substituting $F(\rho)=\sqrt{\rho} f(\rho)$ in the 2D effective radial equation of the $N$-body problem at unitarity [Eq.\eqref{eq:Schrodi_hyper_radial} in the main text] gives
\begin{equation}
-\frac{\hbar^2}{2m_{\rm r}} \left(\partial_\rho^2 +\frac{2}{\rho} \partial_\rho \right) f(\rho) 
+ \frac{\hbar^2 \left(\expos^2-\frac{1}{4}\right)}{2m_{\rm r}\rho^2} f(\rho)  =Ef(\rho) 
\label{eq:Schrodi_3D_radial}
\end{equation}
and thus the 3D mapping is obtained with
\begin{equation}
\expos = l+ \frac{1}{2} .
\end{equation}
In the vicinity of a resonance in the $l$-wave, the partial scattering amplitude in the $l$-wave is of the form
\begin{equation}
f_l(k)=\frac{-k^{2l}}{\frac{1}{w_l} +\alpha_l k^2 + \dots+ i k^{2l+1}} .
\label{eq:partial_wave}
\end{equation}
In the effective range approximation, the terms  'in the dots' of the denominator in Eq.~\eqref{eq:partial_wave} are neglected as was done in the contact model of Refs.~\cite{Pri06a,Pri06b}. The two scattering parameters $w_l$ and  $\alpha_l$ generalize the notion of scattering length and effective range in the s wave scattering. The bound and quasi-bound states correspond to the pole of the denominator and thus verify at the lowest order:
\begin{equation}
\frac{1}{w_l} +\alpha_l k^2 + i k^{2l+1} =0 .
\label{eq:pole}
\end{equation}
At $\expos=1/2$, the 3D mapping gives the s wave resonant problem $l=0$. The effective range $\alpha_0$ is negligible with respect to the scattering length ${w_0=-\Radius/\epsilon}$ and can be neglected (it is exactly zero if one considers the log-derivative condition without any further approximation, see Ref.~\cite{s=.5} in the main paper). There is no quasi-bound state and the pole of the scattering amplitude is at the binding wave number $q=-i k=1/a$, compatible with a bound state for $\epsilon<0$ or $w_0>0$. In a high partial wave ($l\ge 1$), for ${w_l>0}$ (${\epsilon<0}$), there is a shallow bound state of binding energy
\begin{equation}
E=-\frac{\hbar^2}{2m_r w_l \alpha_l}
\end{equation}
and for ${w_l<0}$, there is a low energy quasi-bound state of energy ${E_{\rm r}=\frac{\hbar^2}{2m_{\rm r}}k_{\rm r}^2}$ and width ${\Gamma}$ given by  
\begin{equation}
E_{\rm r}=-\frac{\hbar^2}{2m_r w_l \alpha_l} \quad , \quad \frac{\Gamma}{2E_{\rm r}} = \frac{k_{\rm r}^{2l-1}}{\alpha_l} .
\end{equation}
The width of the resonance, is inversely proportional to  $\alpha_l$, which can be thus denoted 'width parameter' rather than 'generalized effective range' because it does not have the dimension of a length.

In the limit of low energies, there is an equivalence between the contact model of the $N$-body resonance and the effective range approximation used in Eq.~\eqref{eq:pole}. Keeping only the two first terms in the expression of the $N$-body contact condition gives the approximate equation used to obtain the (quasi-) bound state energy of the $N$-body resonance at large $\expos$. By identifying this last equation with Eq.~\eqref{eq:pole}, one finds:
\begin{equation}
\frac{1}{w_l} = - \frac{\epsilon }{R^{2l+1}} [(2l-1)!!]^2 \quad , \quad  \alpha_l = R^{1-2l} (2l-1)!! (2l-3)!!.
\end{equation}
with the convention ${(-1)!!=1}$ for $l=1$. Using the notation $\Rsep$ for the minimal radius where the contact state coincides with the reference state, the 'width-radius' inequality derived from the scalar product of Ref.~\cite{Pri06b} [see Eq.~(64) in this latter reference] can be written 
\begin{equation}
\alpha_l \Rsep^{2l-1} \gtrsim (2l-1)!! (2l-3)!!
\end{equation}
which coincides exactly with the 'width-radius' inequality in the main text of the paper  i.e. ${ \Radius<\Rsep}$. 

\end{widetext}


\begin{thebibliography}{9999}
\bibitem{nodal} A nodal condition at some finite hyper-radius excludes the possibility of a bound state as the Macdonald function is always positive for a real value of $\expos$.
\bibitem{Log-deriv} It will appear important in the following (see Eq.~\eqref{eq:radius_inequality}) that the log-derivative condition is considered here for the contact wave function and not the reference function as in Ref.\cite{Efi71}.
\bibitem{s=.5} Remarkably, for $s=\frac{1}{2}$, Eq.~\eqref{eq:log_derivative} admits the simple bound state solution ${q\Radius=-\epsilon}$ for ${\epsilon<0}$. This result is compatible with the 3D mapping ${l=\expos-\frac{1}{2}}$ corresponding to a s wave resonant two-body problem which is a one-parameter model with the two-body scattering length ${a=-\Radius/\epsilon}$.
\bibitem{seriesKs} When $\expos$ is not an integer, in the limit  ${z\to 0}$
$$
K_{\expos} (z) =  \left[\left(\left(\frac{z}{2}\right)^{-\expos}  \sum_{k=0}^{\infty}  
\frac{\Gamma(\expos-k)}{2 k!} \left(\frac{-z^2}{4}\right)^k  \right)+  \expos \leftrightarrow -\expos\right]
$$
\bibitem{s=1} For small and negative values of the detuning $\epsilon<0$, at ${\expos=1}$, the binding energy is ${E\simeq - 2\Erad \epsilon/W_{-1}(x)}$.
\bibitem{Contact_Efimov} It is worth pointing out that in the Efimov regime, the $N$-body contact condition can be written in term of a single parameter $R^\star$ in the form 
$
{{\displaystyle \lim_{\rho\to 0}} \bigg] (R^\star \rho)^{i|\expos|} + (R^\star \rho)^{-i|\expos|} , F(\rho)\bigg[  =0 .
}
$
In this regime, ${F(\rho)}$ exhibits log-periodic oscillations in the separable region. Instead of this contact condition, the phase can be also imposed by using a nodal (and thus a one-parameter) condition to be compared with Eq.~\eqref{eq:log_derivative} which is a two-parameter condition. The nodal condition is ${F(R_n)=0}$ where  ${R_n=R^\star e^{2n\pi/|s|}}$ and $n$ is chosen for having ${R_n}$ in the separable region \cite{Pri10b}.
\bibitem{seriesKn} When $\expos=n>0$ is an integer, one has in the limit  $z\to 0$ 
$$
K_n(z)= \left(\frac{z}{2}\right)^{-n}
\sum_{k=0}^{n-1} \frac{\Gamma(n-k)}{2k!} \left(\frac{-z^2}{4}\right)^k - \left(\frac{-z}{2}\right)^n \frac{\ln(z c_n)}{ n!}  + \dots
$$
with $c_n=1/2\times \exp(\gamma_E-(1+\dots1/n)/2)$ and $\gamma_E$ is the Euler's constant:
${c_0=\frac{1}{2} e^{\gamma_E}}$, ${c_1=\frac{1}{2} e^{\gamma_E-1/2}}$, ${c_2=\frac{1}{2} e^{\gamma_E-3/4}}$, \dots
\bibitem{s=0} For ${\expos=0}$, the  contact condition is given by imposing the behavior ${F(\rho) = A\ln (\rho/a_0)+O(\rho^2)}$ as ${\rho\to 0}$ where ${a_0=\Radius \exp(-1/\epsilon)}$ is analogous to a two-dimensional scattering length. An alternative way to impose the contact condition is to use the pseudo-potential for a two-dimensional resonant s wave interaction with the change ${a_{\rm 2D} \to a_0}$ \cite{Ols01}.
\bibitem{large_s} The index $\expos$ is an increasing function of the number $N$ of interacting particles with the asymptotic law ${\expos\propto N^{4/3}}$ (see page 127 in \cite{Wer08}).
\end{thebibliography}
\end{document}